# Low-Dose CT Image Enhancement Using Deep Learning

Ahmet Demir*[1], Mohamad M.A. Shames[1], Omer N. Gerek[2], Semih Ergin[1], Mehmet Fidan[3], Mehmet Koc[4], M. Bilginer Gulmezoglu[1], Atalay Barkana[2], and Cuneyt Calisir[5]

[1]Department of Electrical and Electronics Engineering, Eskisehir Osmangazi University
[2]Department of Electrical and Electronics Engineering, Eskisehir Technical University
[3]Vocational School of Transportation, Eskisehir Technical University
[4]Department of Computer Engineering, Eskisehir Technical University
[5]Department of Radiology, Eskisehir Osmangazi University

January 10, 2023

**Abstract**

The application of ionizing radiation for diagnostic imaging is common around the globe. However, the process of imaging, itself, remains to be a relatively hazardous operation. Therefore, it is preferable to use as low a dose of ionizing radiation as possible, particularly in computed tomography (CT) imaging systems, where multiple x-ray operations are performed for the reconstruction of slices of body tissues. A popular method for radiation dose reduction in CT imaging is known as the quarter-dose technique, which reduces the x-ray dose but can cause a loss of image sharpness. Since CT image reconstruction from directional x-rays is a nonlinear process, it is analytically difficult to correct the effect of dose reduction on image quality. Recent and popular deep-learning approaches provide an intriguing possibility of image enhancement for low-dose artifacts. Some recent works propose combinations of multiple deep-learning and classical methods for this purpose, which over-complicate the process. However, it is observed here that the straight utilization of the well-known U-NET provides very successful results for the correction of low-dose artifacts. Blind tests with actual radiologists reveal that the U-NET enhanced quarter-dose CT images not only provide an immense visual improvement over the low-dose versions, but also become diagnostically preferable images, even when compared to their full-dose CT versions.

## 1 Introduction

The human body organs and extremities are visualized using medical imaging methods to identify several diseases [1]. Medical imaging is an important tool used in medical and biological research, and imaging systems often use an image reconstruction algorithm to create final visualizations of images [2]. The concept of image reconstruction is an inverse mathematical operation for mapping the sensor domain information to the image domain. A good image reconstruction is a key component for establishing high-quality images from sensors [3]. The goal of medical image reconstruction is to get high-quality medical images for clinical use at the lowest possible cost and risk to the patients [1]. One of the important and commonly used imaging methods is Computed Tomography (CT), where multiple x-ray images that are acquired at multiple orientations orthogonal to the slice to be visualized are used for reconstructing an image using modern or classical back projection methods [4, 5]. Since CT corresponds to an indirect tissue slice imaging, it offers information regarding vascular networks, luminal patency, and spatial geometry [6]. However, due to multiple X-ray operations, a patient receives high doses of radiation. Furthermore, some imaging contrast-boosting chemicals may aggravate situations with kidney diseases [7]. Low-dose CT is gaining popularity as a means of lower radiation exposure. However, a direct reduction of the radiation dose results in a considerable degradation in image quality [8]. As part of a rapidly growing discipline, deep learning (DL) and other machine learning (ML) approaches provide a promising method for image reconstruction, with artifact reduction and reconstruction speed-up [8, 9, 10, 11, 12]. Deep neural network (DNN) based image enhancement in medical imaging has shown encouraging results in under-sampled and low-dose settings. However, such methods normally need a massive amount of data for the training, which require a large amount of computer memory and process time [13]. Our goal in this study is to reduce the distortions caused by low-dose CT scanning using a pre-trained autoencoder-type DNN. Since the concept of distortion, together with its elimination, is a subjective issue that depends on the image content, biomedical images must be carefully handled and evaluated according to their performance for direct and correct diagnosis. In this study, we have incorporated practicing radiologists into the research group and performed blind opinion surveys to score how well distortion elimination via DNN works regarding diagnosis-wise visual information availability.

---

*ahmet.demir@ogu.edu.tr

## 1.1 Literature Review

It is seen that image enhancement methods based on deep learning in the literature contribute positively to both qualitative and quantitative improvement of images. On the other hand, most deep learning techniques show computational complexity, requiring large training data sets. Besides, they are difficult to interpret, explain, and generalize. The majority of deep learning-based studies in the literature use open-source imaging data sets available for medical image processing. These studies have focused on open-science medical imaging research, including open-source software packages. Although related articles include a large number of general image processing applications that describe the specific deep learning technique and its application in detail, few examine deep learning applications in medical image enhancement [1].

Image enhancement in low-dose CT or CT with limited angle problems was attempted in a recent study with a novel neural network for 2D sparse image [13]. A tutorial by McCann et al. introduces the basic concepts of systems modeling and biomedical image enhancement methods using modern sparsity and learning-based approaches [2]. The tutorial explains how the system model to be used for describing a wide variety of imaging modalities can be created by integrating several blocks. In addition, image enhancement algorithms are discussed by grouping them into three general categories. The first category includes conventional and direct methods, including the Tikhonov arrangement; the second consists of sparsity and compression/detection theory-based methods; and the third category consists of learning-based (data-driven) methods, including various DNNs.

Yedder et al. comparatively examined the basic image enhancement algorithms used in the literature and state-of-the-art image enhancement algorithms based on deep learning-based methods in terms of applied measurements, datasets, and key challenges to propose potentially strategic directions for new studies [9].

Reader et al. introduced traditional PET image development methods and then explained the principles of general back projection mapping from measurements to images. In addition, they discussed nonlinear problems that can be used in convolutional DNNs [14]. A deep learning-based back-projection methodology for PET image enhancement was reviewed. Such methods reportedly learn the view-port and data statistics from scratch without relying on a priori knowledge of these data patterns. In contrast, model-based or physics-informed deep learning reportedly uses back projection tools in PET image development and replaces traditional components with data-driven deep learning counterparts such as regularization. These methods rely on statistics from training data samples to learn deep maps for reconstruction while using reliable models of real imaging physics and noise distribution [14].

Haan et al. introduce the use of deep learning for optical sensing systems and computational microscopy [15]. The study reviews the fundamentals of inverse problems in optical microscopy and outlines deep learning methods to solve these problems with supervised methods. It also discusses deep learning applications for image enhancement and getting super-resolution from single images [15].

Chen et al. use a three-layer shallow convolutional neural network to remove noise on low-dose CT images by learning a feature mapping from low- to the corresponding normal-dose images [8]. The architecture splits the noisy CT images into image patches, then denoises the patches before reconstructing a new CT image. They compare their architecture with various state-of-the-art methods with regard to PSNR, RMSE, and SSIM. Chen et al. developed a CNN model that includes a deconvolution network for reducing noise artifacts in low-dose CT images and shortcut connections [17]. This model is called a residual encoder-decoder convolutional neural network (RED-CNN) [10]. The method uses an encoder-decoder network which is similar to U-NET. Kang et al. followed a similar method, but they used directional wavelet transform of CT images [12]. The wavelet network adopted the shortcut connections included in the U-NET [16] directly, and the RED-CNN replaced the pooling/unpooling layers of the U-NET with convolution/deconvolution pairs. Generative adversarial networks are first applied to noise reduction problems in CT images by Wolterink et al. [11]. In their work, the first (generative) network generates a high-dose CT image from a low-dose, which is the noisy one, while the second (adversarial) network decides if the generated image is realistic.

The image enhancement literature is not limited to low-dose CT images, and several deep-learning methods were proposed and applied for the enhancement of various types of images [18, 19, 20]. A common practice is to alter available DNN architectures or combine information from various network channels. For example, the techniques described in [10, 11, 12] involve auto-encoders, and, therefore, they are closely related to our methodology which also incorporates a U-NET autoencoder. However, they all propose adaptations of the architectures with more complex and computationally expensive operations, such as patch encoding and wavelet transform computations. Besides, although the enhancements should address the diagnosis requirements of actual radiologists, these studies lack blind tests with real radiologists for diagnostic accuracy after the enhancements. In our study, we have observed that the direct employment of U-NET readily generates images with reduced noise and retained sharpness, indicating applicability for diagnostic accuracy. In order to check the later applicability issue, the results are then verified by two radiologists, allowing us to provide a measure regarding the real usefulness of the proposed enhancement methodology.

## 2 Methodology

U-NET is an autoencoder-type deep network architecture that comprises a contracting (input-side) and an expansive (output-side) path with extra connections between these paths. It was originally developed for the segmentation of biomedical images [16].

In our study, we use the U-NET structure that is given in Figure 1. The figure shows a trained U-NET model, illustrating its input and output images. The input and output image size of U-NET is selected as 256×256 in accordance with the utilized CT images. The contracting path of the network consists of repeated 3×3 convolution layers, each followed by a rectified linear unit (ReLU) layer and a 2×2 max pooling layer with a stride of 2 to double the channel size. The convolution layers extract useful information and create sparse spatial interactions. No padding is used in the contracting path convolutions as it alters the image size. The choice of ReLU in activation layers is to solve the vanishing gradient problem. Max-pooling is used to increase the robustness of the representation to small spatial translations in the input image. At the end of the contraction path, 64 features are collected in one large feature by a 1×1 convolution layer. The second half of the network, named the expansive path, contains convolutional and average pooling layers which restore the output size to the original input image size. As a distinct property of U-NET, certain contracting path features are transferred to the expanding path and concatenated with some expanding path features for more precise localization.

The constructed U-NET is, then, trained from scratch, using 5926 image pairs that contain both full-dose and quarter-dose CT images, taken from the publicly available Low Dose CT Grand Challenge dataset [21]. The dataset contains slices of chest CT images for ten different patients. The original dataset contains images that are taken in two different slice thicknesses of 1 mm and 3 mm, whereas we use the images having a slice thickness of 1 mm.

The aim of the training section is to adapt the U-NET parameters in such a way that a quarter-dose CT image input could yield a network output as close to the corresponding full-dose CT image, as possible. The loss function to measure the similarity between the generated and the desired image is classically selected as the Mean Squared Error (MSE). A total of 5926 image pairs with an image size of 256×256 pixels and with 32 bits/pixel bit depth were used in the training phase. For images with size mismatches, a center-crop operation was performed. Randomly selected 10 % of the images were used for the validation phase in every epoch. The training epoch number was set to 100, and the batch size was set to 4. The learning rate of the used RMSprop optimizer function was set to 0.0001. The training loss function values (i.e., the MSE values) were monitored throughout the training epochs. Figure 2 shows how the MSE value changes over 100 epochs during an actual training phase. An initial MSE value of over 0.0045 value gradually decreases down to 0.0020. It can be seen that the decrease rate of MSE slows down at each epoch, and reaches a nearly steady state after epoch 25, indicating that the selected epoch count is sufficient for the training process.

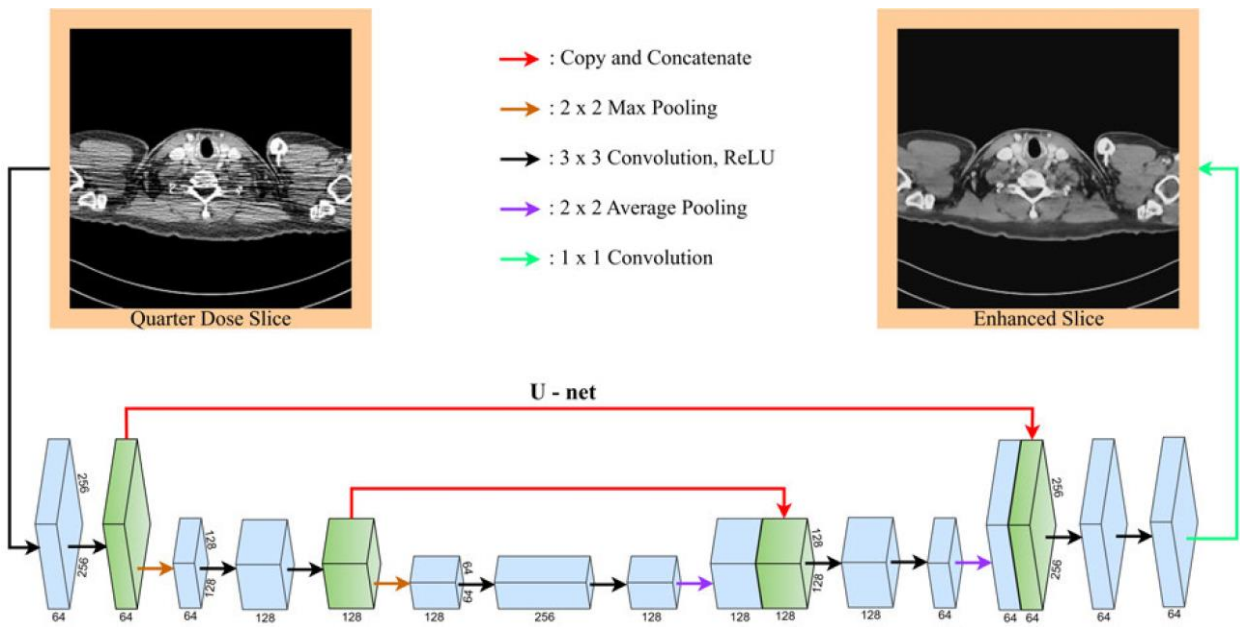

Figure 1: U-net structure

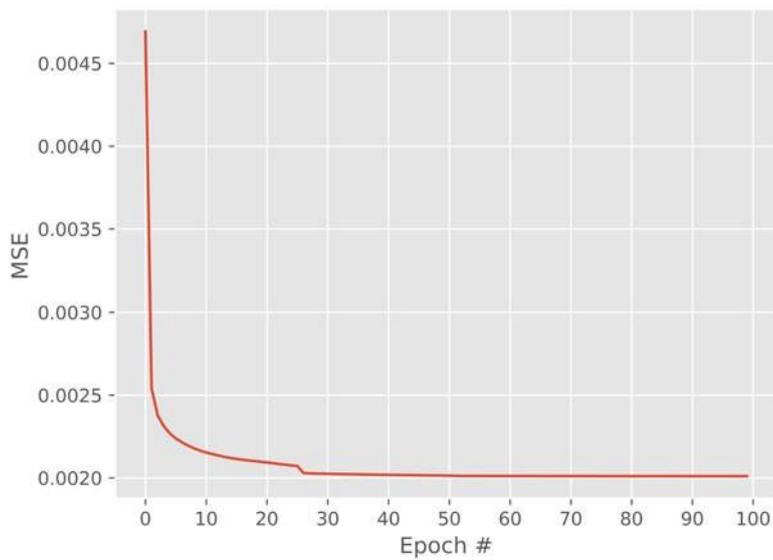

Figure 2: Change of mean squared error with progressing epochs

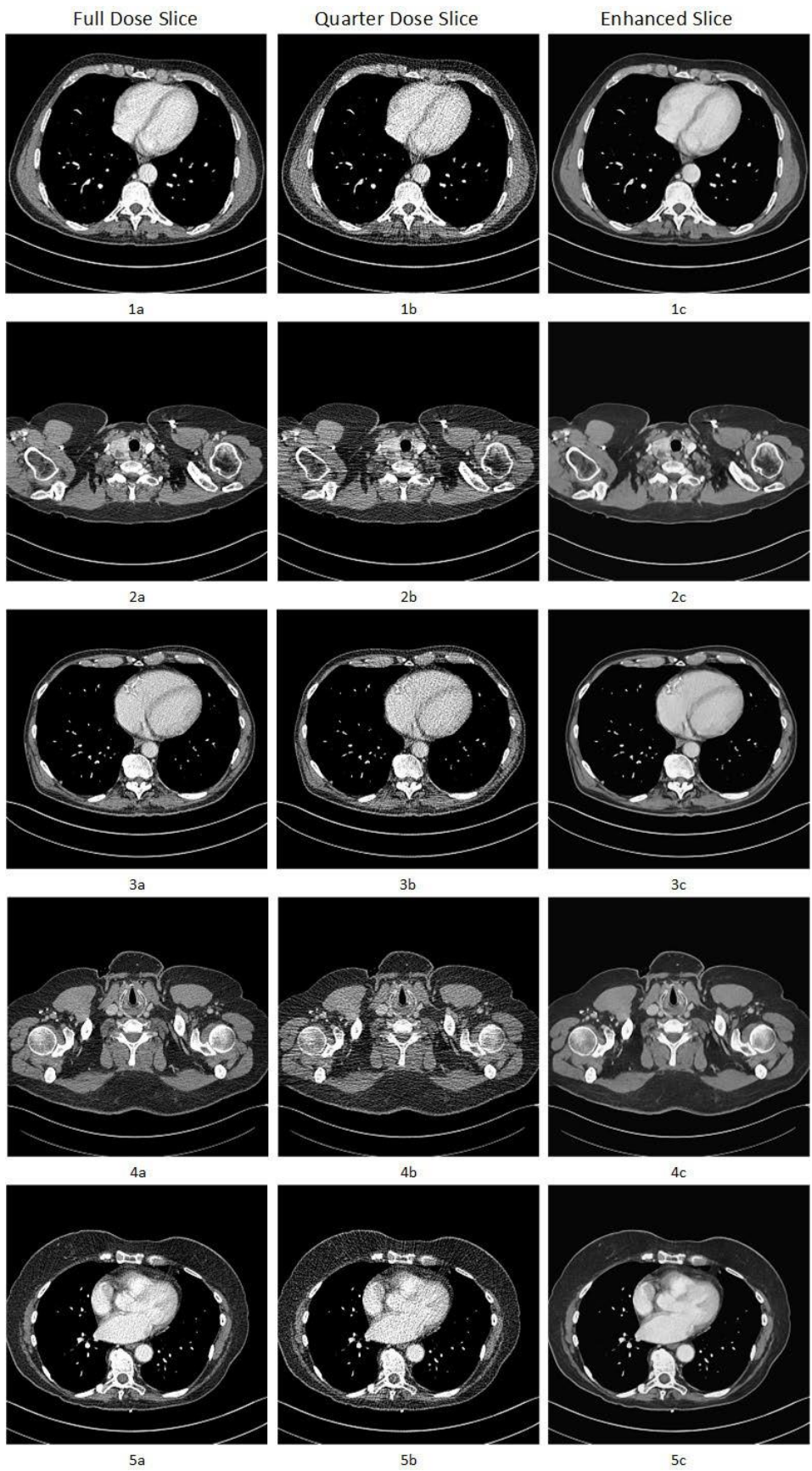

Figure 3: CT slice versions from the patients $P_1$, $P_2$, $P_3$, $P_4$, $P_5$

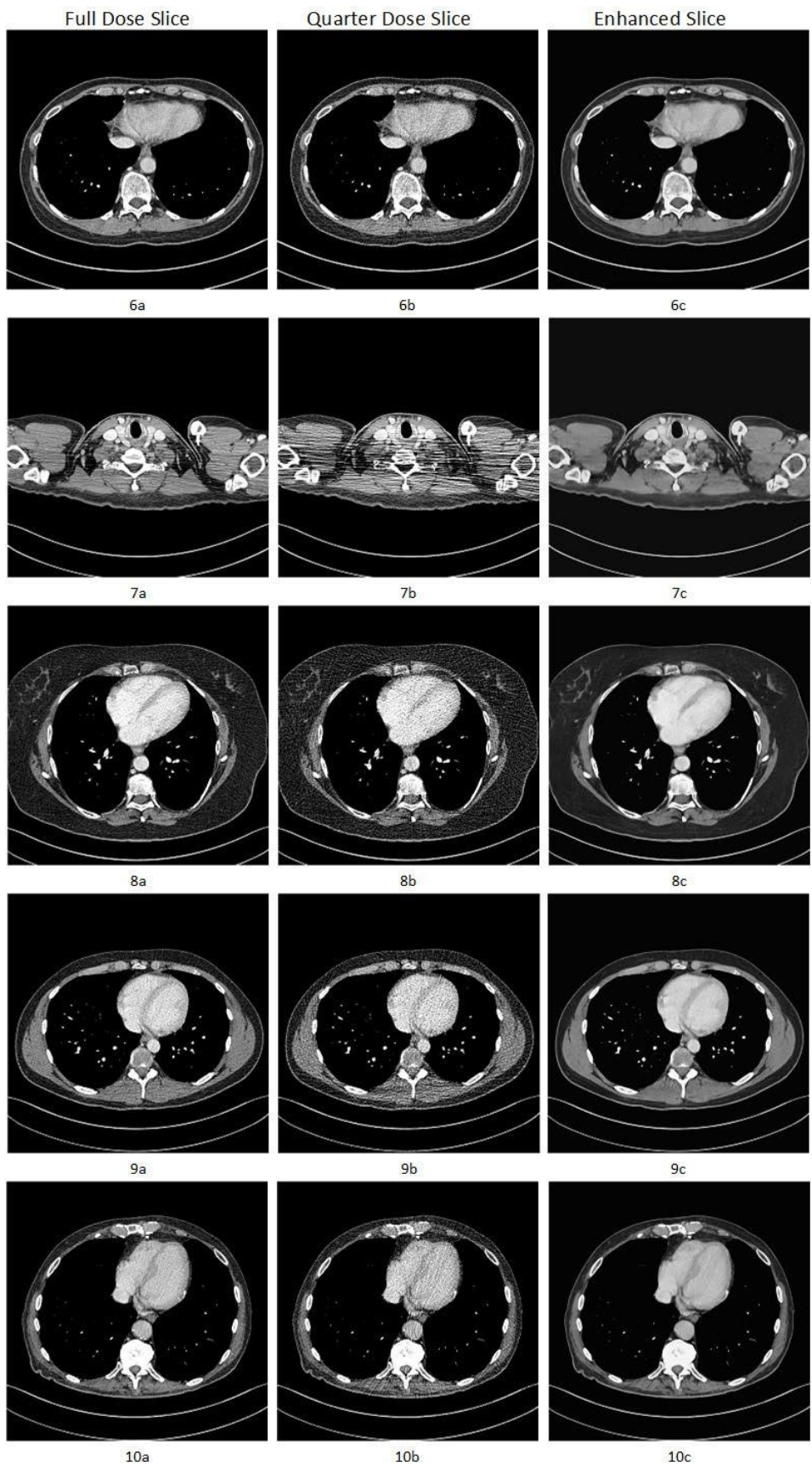

Figure 4: CT slice versions from the patients $P_6$, $P_7$, $P_8$, $P_9$, $P_{10}$

# 3 Experimental Results

Following the training phase that was described in the previous section, a test suit was applied using ∕ 10 distinct CT images, which were also obtained at quarter- and full doses. These quarter-dose images were fed to the trained network, and the resultant images were recorded for quantitative, qualitative, and comparative analysis. Figures 3 and 4 show the results obtained when the model acquired at the end of 100 epoch training is tested on the above-mentioned 10 quarter-dose test images, which were strictly excluded from the training phase. In order to provide visual comparisons, the full dose, quarter-dose, and U-NET-enhanced slices are given in three separate columns for a total of ten different images. It can be seen that the enhancement of the trained U-NET mostly corresponds to the reduction of noisy surface patterns due to CT reconstruction from low-dose angular recordings, whilst retaining even the most subtle edge transitions.

An important aspect of the enhancement is its usefulness according to actual radiologists. Even if a CT image version may look smoother or with better detail, it is eventually up to the radiologist to make a solid evaluation and comparison. It must be noted that the most critical evaluation parameter for the radiologist is the quality of the CT image for correct and accurate diagnosis. A radiologist strictly requires the image to reveal visual hints for pathology detection and classification; therefore, most alterations, although they might look pleasing to the untrained eye, may actually distort the image or occlude important visual hints. In order to consider this perilous possibility, a thorough assessment was carried out using two expert medical radiologists. The full and quarter-dose images, as well as the enhanced images for each CT, are given to the radiologists in random order, and the doctors were asked to quantify the suitability of each image with a subjective opinion score between 1 and 10 (1 being worst, 10 being best). The attained scores from these two independent radiologists are presented in Table 1. The scores reveal that both radiologists overwhelmingly prefer the U-NET-enhanced images over the quarter-dose, meaning that the U-NET enhancement does not deprive the output of necessary visual signatures for diagnostic purposes during the enhancement process. The perplexing observation is that the scores of the enhanced slice versions are even higher than the full-dose versions. Among these comparisons for 10 images, there is only one image case (P10), where one radiologist prefers the full dose over the enhanced version (while the other radiologist still prefers the enhanced version, by a greater vote margin).

In an attempt to quantify the amount of change in the contained information, we have conducted visual correlation experiments. The overlap ratio of the information carried by two separate images can be measured using morphological correlation measurement techniques. Among several metrics for measuring these correlations, a universally accepted metric is the Pearson correlation coefficient, which is defined as the linear correlation between two data sets (pixel values of two images) [22]. It can be expressed as the ratio between the product of the covariance and the standard deviation of two variables, which provides an output value between -1 and 1. All these operations in the calculation of the Pearson coefficient are explained in Eq. 1. In this work, Pearson correlations between Full Dose - Quarter Dose, Full Dose - Enhanced, and Quarter Dose - Enhanced slices for ten selected patients are evaluated and given in Figure 5.

Table 1: On a scale of 1 (worst) - 10 (best), the scores of two radiologists to the full dose, quarter dose, and enhanced slices of 10 patients.

| Patient | Full Dose Slice | | Quarter Dose Slice | | Enhanced Slice | |
|---|---|---|---|---|---|---|
| | Radiologist 1 | Radiologist 2 | Radiologist 1 | Radiologist 2 | Radiologist 1 | Radiologist 2 |
| $P_1$ | 6 | 8 | 3 | 4 | 8 | 9 |
| $P_2$ | 7 | 6 | 4 | 2 | 8 | 6 |
| $P_3$ | 7 | 8 | 5 | 6 | 8 | 8 |
| $P_4$ | 7 | 9 | 5 | 3 | 8 | 9 |
| $P_5$ | 8 | 7 | 5 | 4 | 9 | 8 |
| $P_6$ | 8 | 7 | 6 | 6 | 8 | 7 |
| $P_7$ | 6 | 6 | 2 | 1 | 8 | 8 |
| $P_8$ | 7 | 8 | 4 | 4 | 8 | 9 |
| $P_9$ | 8 | 8 | 5 | 5 | 9 | 9 |
| $P_{10}$ | 8 | 6 | 4 | 4 | 7 | 8 |

$$r = \frac{(x_i - \bar{x})(y_i - \bar{y})}{\sqrt{(x_i - \bar{x})^2 \ (y_i - \bar{y})^2}} \tag{1}$$

$r$ = Pearson correlation coefficient
$x_i$ = values of the x-variable in a sample
$\bar{x}$ = mean of the values of the x-variable
$y_i$ = values of the y-variable in a sample
$\bar{y}$ = mean of the values of the y-variable

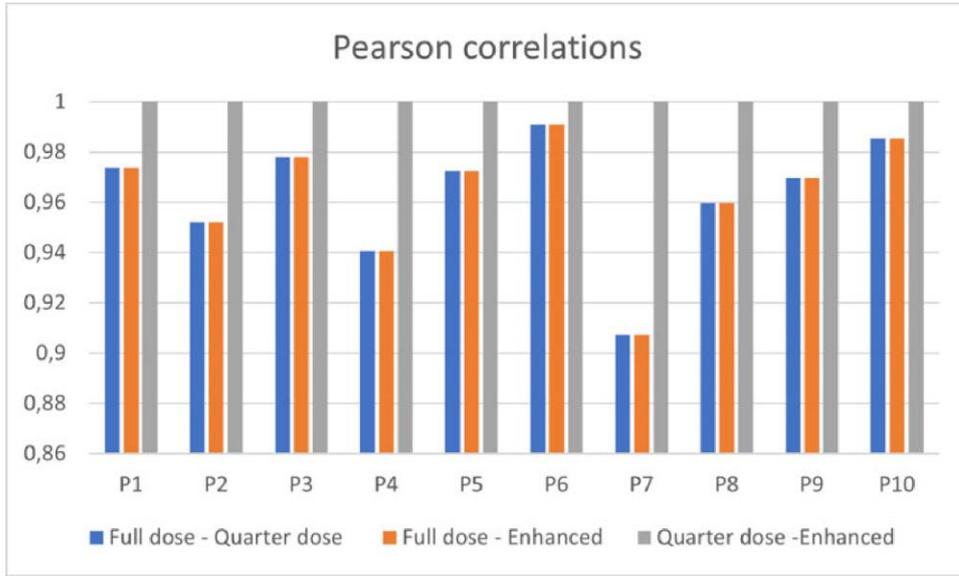

Figure 5: Pearson correlations between pairs of Full dose, Quarter dose, and Enhanced slices

Since covariance can only reflect a linear correlation of variables, it may lack indication of other types of data relationships. In statistics, Spearman's rank correlation coefficient is a non-parametric measure of rank correlation (i.e., statistical dependence between the rank of two variables) [23]. It measures the success of defining the relationship between two variables with a monotonous function. As a result, the Spearman correlation between the two variables measures the Pearson correlation between the *rank values* of these two variables. The calculation of the Spearman coefficient is explained in Eq. 2. By this definition, it can be stated that while the Pearson correlation evaluates linear relationships, the Spearman correlation evaluates nonlinear relationships through a monotonous mapping. If the data values (pixel values) are not repeated, a perfect Spearman correlation of +1 or -1 occurs when each of the variables can be achieved by a perfectly monotonic function of the other. Spearman correlations between Full Dose - Quarter Dose, Full Dose - Enhanced, and Quarter Dose - Enhanced slices for ten selected patients are calculated and shown in Figure 6.

$$r_s = r_{R(X)R(Y)} = \frac{\text{cov}(R(X), R(Y))}{\sigma_{R(X)} \sigma_{R(Y)}} \tag{2}$$

$r$ = Pearson correlation coefficient
$r_s$ = Spearman correlation coefficient
$R(X)$ = Sample rank of X variable
$R(Y)$ = Sample rank of Y variable
$\text{cov}(R(X), R(Y))$ = Covariance of sample ranks of X and Y
$\sigma_{R(X)}$ = Standart deviation of sample rank of X
$\sigma_{R(Y)}$ = Standart deviation of sample rank of Y

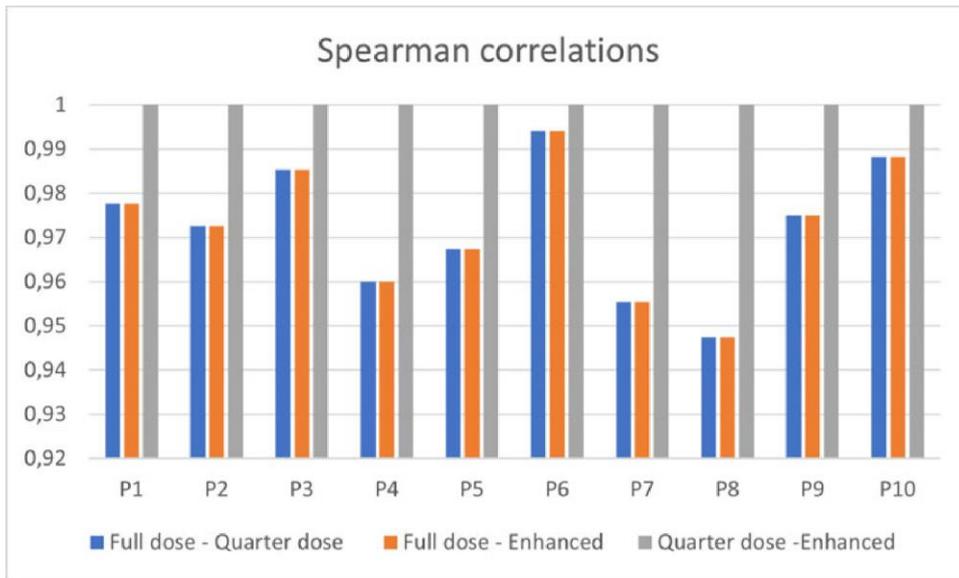

Figure 6: Spearman correlations between pairs of Full dose, Quarter dose and Enhanced slices

An immediate observation from Figure 5 and 6 is that Pearson and Spearman correlations between quarter-dose and enhanced images are identically equal to one. Furthermore, their separate correlations to the full-dose images yield precisely the same correlation values. This observation shows that there is absolutely no morphological distortion or information change in the process of enhancement from the quarter dose. This is a critical observation to prove that the shape information contained in the enhanced image never deteriorates or alters with the enhancement procedure. Besides, no bogus shape information is contaminating the enhanced images. Therefore, the only effect of the performed operation is the increase in the image quality by means of an improved signal-to-noise ratio, which helps radiologists to perform more accurate diagnostic interpretations.

# 4  Discussion and Conclusion

In this paper, it is shown that the enhancement process of U-NET DNN, which was trained for autoencoding using sufficiently many low-dose / full-dose input-output CT image pairs, does not cause any serious data loss for medical imaging. Furthermore, subjective tests show that it allows even better diagnostic assessment on the enhanced CT images. The key aspect of medical assessment is the detection and classification of image patterns, such as nodules or tissue patterns. In radiology, these image elements may degrade due to several factors, such as noise or measurement distortions. These degradations are known as image artifacts, which are also very commonly encountered in clinical computed tomography (CT), and may either degrade image quality and obscure or simulate pathology. Due to the specific image construction methodology of CT, several types of CT artifacts may occur, ranging from additive noise to beam hardening, from irradiance beam scatter to patient motion, and from metal particles in the tissue to low and inaccurate signal fidelity. Since CT is a reconstruction from thin x-ray data that are obtained at various rotational angles, the reconstructed image is very sensitive to high-attenuation internal tissues, such as bones or tissues with iodinated contrast. If the penetrating x-ray does not attain a sufficient magnitude, its attenuation through such tissues causes saturated observation, and its degradation is usually observed in the form of thin and long streaks along the major axis through individual high-attenuation objects. Another visible artifact due to lower dose x-rays reveals in the form of granularity in the images. With such increased granularity noise, certain high-contrast objects, such as bone, may still be visible, but lower-contrast regions due to soft-tissue boundaries may be obscured and difficult to notice or evaluate. A third visible artifact is known as beam hardening and scatter, where dark streaking bands appear between two high-attenuation objects positioned close to each other. Correct and careful elimination of these artifacts may result in improved image quality, thereby increasing diagnostic accuracy. In thorax CT examinations, since the large bone structures are located close to each other in the sections passing through the thoracic apex, streaking bands are detected more frequently in these sections.

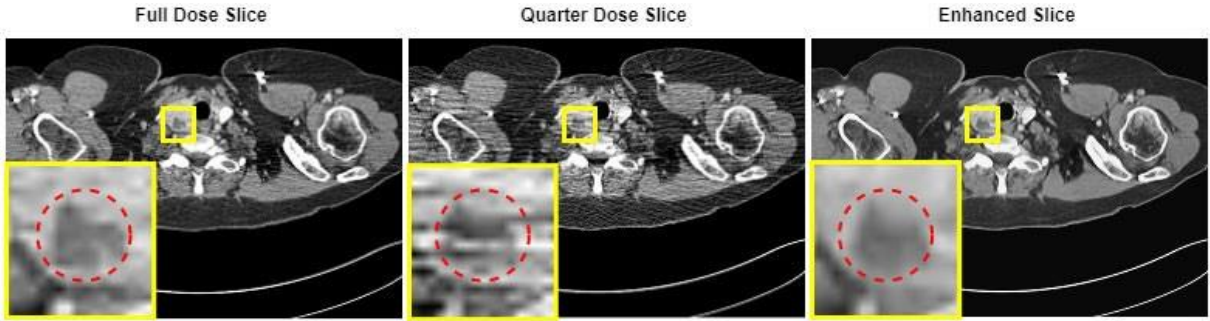

Figure 7: CT slice versions of the patient $P_2$

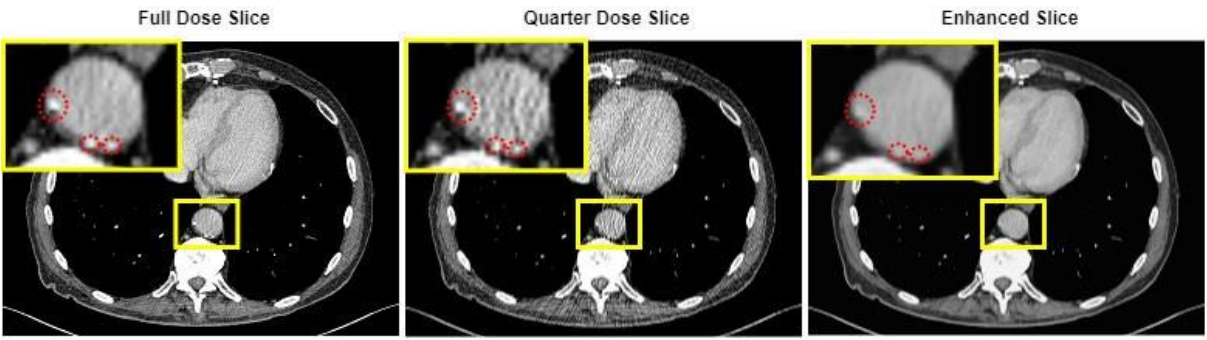

Figure 8: CT slice versions of the patient $P_{10}$

The proposed U-NET system was trained with quarter- and full-dose versions of CT images that were expected to reveal the noise amount difference in the above-mentioned artifact cases. As an example, streaking bands were most frequently observed in the upper thorax sections in the related test images. However, it was observed that the streaking bands present in the quarter-dose images (which were also very prominent even in full-dose images) were significantly reduced in the reconstructed images. In a majority of the radiologist blind tests, the enhanced images were voted to provide better diagnostic quality than even the full-dose images. The areas zoomed in Figure 7 clearly show that the reconstructed image is more informative and rated higher in terms of diagnostic evaluation by both of the experienced radiologists compared to the full dose and quarter dose images. In this image, the radiologists report that the streaking bands of the low- and full-dose CT images considerably complicate the evaluation of the thyroid nodule details, whereas the nodule is better visible in the enhanced image.

Among the test images, there is only one single case, where the full dose image was rated higher than the reconstructed image in terms of diagnostic image quality by two experienced radiologists, while the rest of the reconstructed test images were rated higher, even compared to full dose test images. The individual lower-rated enhancement is given in Figure 8. In this particular case, the reasoning of the lower-voting radiologist was that, while the enhanced image diminished a few mild calcifications in the aortic wall that were of no clinical significance, these calcifications were more clearly visible in both full-dose and quarter-dose images. When such mild calcifications are located within a mass or nodule, they might be of diagnostic importance for lesion classification. Hence, when a nodule is detected in the reconstructed images, we think that it would be more appropriate to decide whether the nodule contains calcification by examining other plane images as well as axial plane images. This evidence indicates that there is always a possibility of data loss in certain regions of the enhanced images, so extreme care must be taken in image alteration processes, including enhancement. As a result, this study not only puts forth a clearly successful and simple U-NET-based enhancement procedure but also indicates the necessity of diagnostic validation and provides clear subjective comparisons among quarter-, full-dose, and enhanced CT images.

# References

[1] E. Ahishakiye, M.B.V. Gijzen, J. Tumwiine, R. Wario, J. Obungoloch, A survey on deep learning in medical image reconstruction, Intelligent Medicine, 1 (3), 2021, pp.118-127.

[2] M.T. McCann, M. Unser, Biomedical Image Reconstruction: From the Foundations to Deep Neural Networks, Foundations and Trends in Signal Processing, 13 (3), 2019, pp. 283-359.

[3] I. Oksuz, J. Clough, A. Bustin, G. Cruz, C. Prieto, R. Botnar, D. Rueckert, J. Schnabel, A. King, Cardiac MR Motion Artefact Correction from K-space Using Deep Learning-Based Reconstruction, International Workshop on Machine Learning for Medical Image Reconstruction, 2018, pp. 21-29.

[4] S.K. Lakshmanaprabu, S.N. Mohanty, K. Shankar, N. Arunkumar, G. Ramirez, Optimal deep learning model for classification of lung cancer on CT images, Future Generat. Comput. Syst. 92, 2019, pp. 374–382.

[5] D. Racine, F. Becce, A. Viry, P. Monnin, B. Thomsen, F.R. Verdun, D.C. Rotzinger, Task-based characterization of a deep learning image reconstruction and comparison with filtered back-projection and a partial model-based iterative reconstruction in abdominal CT: A phantom study, Physica Medica, 76, 2020, pp. 28-37.

[6] J.B. Lin, E.H. Phillips, T.E. Riggins, G.S. Sangha, S. Chakraborty, J.Y. Lee, R.J. Lycke, C.L. Hernandez, A.H. Soepriatna, B.R. Thorne, et al. Imaging of small animal peripheral artery disease models: recent advancements and translational potential, Int J Mol Sci., 16 (5), 2015, 11131–77.

[7] R.Y. Kwong, E.K. Yucel, Computed tomography scan and magnetic resonance imaging, Circulation, 108, 2003, e104–e106.

[8] H. Chen, Y. Zhang, W. Zhang et al, Low-dose CT via convolutional neural network, Biomed Opt Express, 8, 2017, pp. 679–694.

[9] H. B. Yedder, B. Cardoen, G. Hamarneh, Deep learning for biomedical image reconstruction: A survey, Artificial intelligence review 54 (1), 2021, pp. 215-251.

[10] H. Chen et al., Low-dose CT with a residual encoder-decoder convolutional neural network, IEEE Trans. Image Process., 36 (12), 2017, pp. 2524–2535.

[11] J.M. Wolterink, T. Leiner, MA. Viergever, I. Isgum, Generative adversarial networks for noise reduction in low-dose CT, IEEE Trans. Med. Imag., 36 (12), 2017, pp. 2536–2545.

[12] E. Kang, J. Min, and J. C. Ye. (2016). A deep convolutional neural network using directional wavelets for low-dose X-ray CT reconstruction. Available: https://arxiv.org/abs/1610.09736.

[13] D. Wu, K. Kim, Q. Li, Computationally efficient deep neural network for computed tomography image reconstruction, Medical Physics 46 (11), 2019, pp. 4763-4776.

[14] A. J. Reader, G. Corda, A. Mehranian, C. Costa-Luis, S. Ellis, J. A. Schnabel, Deep Learning for PET Image Reconstruction, IEEE Transactions on Radiation and Plasma Medical Sciences, 5 (1), 2021, pp. 1-25.

[15] K. Haan, Y. Rivenson, Y. Wu, A. Ozcan, Deep-Learning-Based Image Reconstruction and Enhancement in Optical Microscopy, Proceedings of The IEEE, 108 (1), 2020, pp. 30-50.

[16] O. Ronneberger, P. Fischer, and T. Brox, U-net: Convolutional networks for biomedical image segmentation, Proc. Int. Conf. Med. Image Comput. Comput.-Assist. Intervent., 2015, pp. 234–241.

[17] M. D. Zeiler, G. W. Taylor, and R. Fergus, Adaptive deconvolutional networks for mid and high level feature learning, Proc. IEEE Int. Conf. Comput. Vis. (ICCV), 2011, pp. 2018–2025.

[18] J. Schlemper, J. Caballero, J. V. Hajnal, A.N. Price and D. Rueckert, A Deep Cascade of Convolutional Neural Networks for Dynamic MR Image Reconstruction, IEEE Transactions on Medical Imaging, 37 (2), 2018, pp. 491-503.

[19] H. Jeelani, et al. Image quality affects deep learning reconstruction of MRI, IEEE 15th International Symposium on Biomedical Imaging, 2018, pp. 357-360.

[20] Z. Ramzi, P. Ciuciu, J.-L. Starck, Benchmarking MRI Reconstruction Neural Networks on Large Public Datasets. Appl. Sci. 10 (5), 2020, 1816.

[21] C. McCollough, Low Dose CT Grand Challenge, the Mayo Clinic, the American Association of Physicists in Medicine, and grants EB017095 and grants EB017185 from the National Institute of Biomedical Imaging and Bioengineering, 2016.

[22] S. Glen, Correlation coefficient: Simple definition, formula, easy steps. StatisticsHowTo. com. Available online: https://www.statisticshowto.com/probability-and-statistics/correlation-coefficient-formula/, 2021.

[23] G.U. Yule, M.G. Kendall, An Introduction to the Theory of Statistics (14th Ed.), Charles Griffin & Co., 1968[1950], pp. 268.